\documentclass[12pt,preprint]{aastex}
\shorttitle{Wind Accretion and Evolution of LS~5039}
\shortauthors{McSwain \& Gies}

\begin{document}

\received{}
\accepted{}

\title{Wind Accretion and Binary Evolution of the Microquasar LS~5039}

\author{M. V. McSwain\altaffilmark{1} and
        D. R. Gies\altaffilmark{1}}

\affil{Center for High Angular Resolution Astronomy\\
Department of Physics and Astronomy \\
Georgia State University, Atlanta, GA  30303\\
Electronic mail: mcswain@chara.gsu.edu, gies@chara.gsu.edu}

\altaffiltext{1}{Visiting Astronomer, Kitt Peak National Observatory,
National Optical Astronomy Observatories, operated by the Association
of Universities for Research in Astronomy, Inc., under contract with
the National Science Foundation.}

\slugcomment{submitted to ApJL}
\paperid{}

%%%%%%%%%%%%%%%%%%%%%%%%%%%%%%%%%%%%%%%%%%%%%%%%%%%%%%%%%%%%%%%

\begin{abstract}
There is much evidence to suggest that stellar wind capture, rather than
Roche lobe overflow, serves as the accretion mechanism onto the compact
secondary object in the massive X-ray binary LS~5039.  The lack of
significant emission combined with only a modest X-ray flux provide
observational evidence that no large-scale mass transfer is occurring
(consistent with our estimate of the 
radius of the O6.5~V((f)) optical star that is smaller than 
its critical Roche radius).  Here we determine the mass loss rate of the
optical star from the broad, residual emission in the H$\alpha$ profile.
Using a stellar wind accretion model for a range in assumed primary
mass, we compute the predicted X-ray luminosity for the system.  We
compare our results to the observed X-ray luminosity to determine the mass
of the compact object for each case.  The companion appears to be a
neutron star with a mass between 1 and $3 M_\odot$.  With our new
constraints on the masses of both components, we discuss their
implications on the evolution of the system before and after the supernova
event that created the compact companion.  The binary experienced significant mass loss
during the supernova, and we find that the predictions for the resulting runaway velocity
agree well with the observed peculiar space velocity.  
LS~5039 may be the fastest runaway object among known massive
X-ray binaries.
\end{abstract}


%%%%%%%%%%%%%%%%%%%%%%%%%%%%%%%%%%%%%%%%%%%%%%%%%%%%%%%%%%%%%%%

\section{Introduction}                              % Section 1

LS~5039 is a relatively faint ($V = 11.3$) and massive star of 
type O6.5~V((f)) \citep{cla01} that is
one of only a few confirmed massive X-ray binaries (MXRBs)
with associated radio emission \citep{rib99}.
It has radio-emitting relativistic
jets characteristic of galactic microquasars, and
it is probably a high energy gamma ray source as well \citep{par00}.
We recently discovered that the system is a short period binary ($P = 4.117 \pm
0.011$ d) with the highest known eccentricity ($e = 0.41\pm0.05$)
among O star binaries with comparable periods \citep{mcs01a}.
This high eccentricity probably results from the huge mass 
loss that occurred with the supernova (SN) explosion that gave 
birth to the compact star in the system \citep{bha91,nel99}. 
Binaries that suffer large mass loss in a SN are expected 
to become runaway stars, and recently both we \citep{mcs01b} and 
\citet{rib02} found that LS~5039 has a relatively large proper 
motion that indicates that the binary has a record-breaking 
peculiar space velocity among MXRBs. 

Reliable estimates for the masses of the components of the binary 
are of key importance for any discussion about the evolution 
of this remarkable system.  Here we present an investigation 
of the possible mass range for the X-ray star based on 
a stellar wind accretion model for the 
X-ray production (\S2).  This analysis relies on our observations 
of the wind emission effects in the H$\alpha$ profile and the 
wind models of \citet{pul96}.   We then apply our derived 
mass estimates for both stars to determine the probable masses
and orbital parameters prior to the SN based on the system's  
current eccentricity (\S3).   We find that the predicted 
and observed runaway velocities are in good agreement, 
and, thus, LS~5039 provides the best verification to date 
of model predictions about the outcome of a SN explosion 
in a massive binary. 
 
%%%%%%%%%%%%%%%%%%%%%%%%%%%%%%%%%%%%%%%%%%%%%%%%%%%%%%%%%%%%%%%

\section{X-ray Fueling by Wind Accretion}           % Section 2

The X-ray production in MXRBs results from 
mass accretion through a Roche lobe overflow (RLOF) stream or by 
Bondi-Hoyle accretion from the stellar wind of the luminous 
primary \citep{kap98}.  The systems experiencing RLOF 
usually have large X-ray luminosities and striking optical 
emission lines.  However, LS~5039 has a modest X-ray flux 
\citep{rib99} and no obvious emission lines \citep{cla01,mcs01a}, 
and we show below that the O-star is probably smaller 
than its critical Roche surface.  Thus, we can safely assume 
that most of the mass accretion in LS~5039 occurs through capture of 
the primary's stellar wind flow.  

The wind accretion luminosity of MXRBs depends on the 
system separation, wind velocity law, mass loss rate, 
and mass of the accretor \citep{lam76}.  If 
we can estimate these binary and wind parameters from 
spectroscopy, then the mass of the compact star can be 
found by comparing the predicted and observed X-ray 
luminosities, $L_X$.   Here we present such an analysis 
based on the mass loss rate derived from our observations 
of the H$\alpha$ profile \citep{mcs01a}.   All the 
parameters needed in this analysis ultimately depend 
on the assumed mass of the O-star primary that is 
poorly constrained at the moment.   The simplest assumption 
is that the primary has a mass typical of single stars 
of its spectral classification, approximately $40 M_\odot$ 
(see the mass calibration of 
\citet{hp89} and the study of comparable stars in the 
young binary, DH~Cep, by \citet{pen97}).   On the other 
hand, there is evidence that the primaries in 
MXRBs may be undermassive for their luminosity \citep{kap01}, 
and in extreme cases, their mass may be a factor of three 
lower than the mass derived by comparing their position 
in the Hertzsprung-Russell diagram with evolutionary tracks. 
Thus, we show here the results of a wind accretion model 
for LS~5039 based on a range in assumed primary mass of 
10 -- $40 M_\odot$.

\citet{pul96} show how the H$\alpha$ profile in O-stars grows 
from a pure absorption feature to a strong emission line with 
increasing stellar wind mass loss rate, and they present 
a scheme to estimate the mass loss rate based on the observed 
equivalent width, $W_\lambda$, of H$\alpha$.   We observed 
the H$\alpha$ line in the spectrum of LS~5039 over three runs
in 1998, 1999, and 2000 with the Coude Feed Telescope at
Kitt Peak National Observatory (see \citet{mcs01a} for details), 
and we show in Figure~1 the average profile from each run 
after shifting each spectrum to the rest frame and 
convolving the 1998 and 2000 spectra to the resolution of 
the 1999 run ($R=\lambda / \triangle\lambda = 4000$).  
The H$\alpha$ profile appears to be filled in by broad, residual 
emission that appears as shallow emission peaks in the wings
(perhaps also present in the vicinity of 
\ion{He}{1} $\lambda 6678$ and \ion{He}{2} $\lambda 6683$). 
The emission was apparently stronger during the 1998 and 
1999 runs, and we compare in Figure~1 these profiles with the 
deeper absorption observed in 2000 ({\it dotted lines}). 
The net equivalent width over the entire emission and 
absorption blend was 2.22, 2.15, and 3.10 \AA ~($\pm 10\%$) 
for the average profiles from 1998, 1999, and 2000, respectively. 
We use the mean of these values, 2.49 \AA , in the analysis below. 
We see no evidence in our spectra of variations in the 
residual emission with orbital phase, and so we can 
reliably assume that this weak emission forms in the wind 
of the primary (and not, for example, in a gas stream or 
accretion disk in which we would observe orbital 
variations in radial velocity). 

\placefigure{fig1}     % Figure 1 - H-alpha profiles 

We need estimates of the underlying photospheric component
of absorption equivalent width, effective temperature,
$T_{\rm eff}$, radius, $R_O$, and wind velocity law 
in order to derive a mass loss rate from the 
observed equivalent widths using the wind models of
\citet{pul96} (see their Table~7).   We adopted the 
spectral classification calibration of \citet{hp89} 
to estimate $T_{\rm eff} = 40.5\pm2$~kK and 
$\log g = 4.0\pm0.2$ for the O6.5~V((f)) primary, 
which is in reasonable agreement with the 
results from Str\"{o}mgren photometry by 
\citet{kil93} and with the values of these parameters 
derived by \citet{pul96} for similar stars. 
The radius, $R_O$, follows from $\log g$ and the 
assumed mass, $M_O$.  We assumed the wind velocity 
law usually applied to O-stars: 
$v(r) = v_\infty (1 - R_O/r)^\beta$
with $\beta = 0.8$, $v_\infty = 2.6 v_{\rm esc}$, 
$v_{\rm esc} = (2 (1-\Gamma) G M_O / R_O)^{1/2}$, 
and $\Gamma = 2.6\times 10^{-5} (L_O/L_\odot) / (M_O/M_\odot)$
\citep{hp89,pul96,lam99}.   The final parameter to 
be set is the appropriate photospheric absorption 
equivalent width.  Unfortunately, \citet{pul96} only 
give one value of $W_\lambda = 3.29$~\AA ~that is specified 
for a model photosphere with $T_{\rm eff} = 40$~kK 
and $\log g = 3.7$.  While this effective temperature is 
nearly applicable here, we expect the equivalent width 
will be larger in our $\log g = 4.0$ case since the 
Stark broadening of the Balmer lines increases with gravity. 
We used the NLTE models of \citet{aue72} to estimate that
the H$\alpha$ equivalent width should be $10\%$ stronger 
at $\log g = 4.0$ compared to $\log g = 3.7$ 
(for $T_{\rm eff} = 40$~kK), and we pro-rated accordingly the 
photospheric equivalent width given by \citet{pul96}.  
Our derived estimates of $R_O$, $v_\infty$, and 
mass loss rate $\log \dot{M}$ are given for several 
assumed values of $M_O$ in Table~1.  The mass loss 
rate values are close to the average found for O6.5~V 
stars by \citet{hp89}, $\log \dot{M} = -6.3$ 
(in units of $M_\odot$~y$^{-1}$). 

\placetable{tab1}      % Table 1  - Stellar parameters versus M_O

We next calculated the predicted X-ray luminosity for a range in assumed
secondary mass, $M_X$, using the wind accretion model of \citet{lam76} 
(using the wind velocity law noted above).  Their expression
for the X-ray accretion luminosity is 
$$L_X = (3\times 10^{13} L_\odot) ~\zeta S_a M_X / M_\odot$$
where the efficiency factor for the conversion of accreted matter
to X-ray flux is $\zeta \approx 0.1$.  The accretion rate is
$$S_a = \pi r_a^2 \dot{M} / (4\pi a^2)$$
for a separation $a$ and an accretion radius 
$$r_a = 2 G M_X / v_{\rm rel}^2$$ 
where $v_{\rm rel}$ is the flow velocity of the wind relative to 
the accreting star (see their eq.~10a). 
Given the two assumed masses and the known period, 
we determined the time-averaged separation of the binary 
and then calculated $L_X$ using our estimates 
of the wind speed and mass loss rate from above. 
A sample result is shown in Figure~2 for the average 
and extreme values of the derived mass loss rate 
(from the 1999 and 2000 H$\alpha$ equivalent widths) 
for a test primary mass of $M_O = 25 M_\odot$. 

\placefigure{fig2}     % Figure 2 - L_X versus M_X

The observed value of $L_X$ depends on the assumed distance. 
We used the magnitude $V=11.35\pm 0.1$ and reddening $E(B-V)=1.2\pm0.1$
from \citet{cla01}, and then estimated the star's angular diameter 
by comparing the unreddened magnitude with model magnitudes
from \citet{kur94} for the adopted $T_{\rm eff}$ and $\log g$ 
(calibrated with the fundamental $T_{\rm eff}$ 
data of \citet{cod76}).   The distance is then found 
from the working value of the stellar radius, 
$d = 0.32~R_O/ R_\odot$~kpc.  The observed X-ray luminosity 
was adjusted from the measurements of \citet{rib99} who 
assumed a distance of 3.1~kpc.   This prorated estimate
of $L_X$ is shown as the shaded region in Figure~2. 
The estimated mass of the secondary is found where 
the predicted wind accretion luminosity matches the observed $L_X$. 

We repeated the wind accretion calculation for a grid of 
assumed O-star masses, and our final results are 
listed in Table~1 and plotted in a mass diagram in Figure~3. 
The results are relatively insensitive to our assumptions 
about $T_{\rm eff}$ and $\log g$.  We found, for example, 
that models made using $\log g = 3.7$ led to lower mass loss 
rates but also lower wind velocities, so that the wind accretion 
rates changed very little.   Because the slope $d L_X / d M_X$ 
is relatively large (see Fig.~2), the details of the accretion 
model (for example, the value of the efficiency parameter, $\zeta$)
do not greatly affect the implied secondary masses. 
In fact, the largest uncertainty in the results comes from the 
variation in H$\alpha$ equivalent width between observing runs, 
which presumably reflects significant changes in the stellar 
wind mass loss rate.  Table~1 also lists the size of the Roche 
radius at periastron, and we find that the O-star fits 
comfortably within its Roche lobe over the full range in assumed mass.  
Our results indicate that the secondary has a mass between
1 and $3 M_\odot$ (based on the extremes of the observed 
H$\alpha$ equivalent width and adopted primary mass range), 
and, thus, the secondary is probably a neutron star. 

\placefigure{fig3}     % Figure 3 - Mass diagram 

%%%%%%%%%%%%%%%%%%%%%%%%%%%%%%%%%%%%%%%%%%%%%%%%%%%%%%%%%%%%%%%

\section{A Supernova in a Binary}                   % Section 3
  
The catastrophe of a SN explosion in a massive binary has 
two immediate consequences.  First, the system acquires a large eccentricity 
that is directly related to the amount of mass lost in the SN event. 
The periastron separation in the altered orbit corresponds to the 
pre-supernova semi-major axis.  Second, by conservation of momentum, 
we expect the entire system to attain a runaway velocity that again 
depends on the mass lost in the SN.   We demonstrate here that 
the values of the observed eccentricity and the probable masses 
derived above indicate that a huge amount of mass was lost in the 
SN event that formed LS~5039. 

We can use the observed eccentricity to 
relate the pre- and post-SN orbital parameters 
if we make the following reasonable assumptions: 
(1) the pre-SN orbit was circular (almost certainly the case 
for such a short period and evolved system), 
(2) any kick velocity imparted to the remnant core due to asymmetries in
the explosion was relatively small \citep{nel99}, 
(3) the primary suffered only minor ablation of mass in the 
explosion so that its pre- and post-SN mass is the same \citep{fry81}, and 
(4) the system has experienced little or no tidal reduction of 
the orbital eccentricity since the SN event. 
\citet{bha91} and \citet{nel99} give the expressions 
required to calculate the pre-SN parameters, and 
we list in Table~2 the resulting orbital and physical 
parameters for an eccentricity $e=0.41\pm 0.05$ \citep{mcs01a}
and the masses derived in \S2.   Table~2 gives the 
pre-SN period, $P^{\rm initial}$, 
semi-major axis, $a^{\rm initial}$, 
SN precursor mass, $M_2^{\rm initial}$, and 
the mass lost in the SN, $\triangle M_2$. 

\placetable{tab2}      % Table 2  - Pre/Post SN parameters 

The first striking result is the large mass loss that 
occurred in the SN in LS~5039.  Some 5 - $17 M_\odot$ was 
lost in the explosion, amounting to more than 81\% of 
the precursor's mass.   This is much larger than the 
typical inferred mass loss fraction of 35\% found 
by \citet{nel99} for black hole X-ray binaries.  
The SN precursor had a smaller mass than the primary 
at the time of the explosion, so the system was not
in danger of total disruption \citep{nel99}.   

The pre-SN orbit was very compact and the stars were in close proximity. 
We list in Table~2 the sizes of the Roche radii
in the pre-SN stage.  These radii are quite restrictive, 
and we find that the primary overfills the Roche lobe in 
the higher mass solutions.  However, the current radius $R_O$
given in Table~2 may be larger than the radius at the 
time of the SN if the primary has evolved to a larger size 
since then.   Nevertheless, the secondary's Roche radius is 
also quite small, and it is possible that the system was 
in a contact or over-contact configuration at the 
time of the SN.  Some kind of close interaction must have 
occurred since by the time of the SN the mass ratio had 
reversed and the separation had been increasing \citep{wel01}. 
Given this evidence of a pre-SN interaction, the 
observed C deficiency of the primary \citep{mcs01a}
probably results from nuclear-processed gas transferred from the SN progenitor.
The currently faster than synchronous rotation of the primary \citep{mcs01a}
may correspond to the synchronous rate in the shorter period,
pre-SN configuration.

\placetable{tab2}      % Table 2  - Pre/Post SN parameters 

Finally, we list in Table~2 the predicted system runaway velocity, $v_{\rm sys}$,
based on conservation of momentum (with errors propagated from 
the uncertainty in the observed eccentricity).  The predictions 
suggest that the system should have a runaway velocity in excess 
of 100 km~s$^{-1}$.   Although the systemic velocity along the 
line of sight is unexceptional \citep{mcs01a}, the tangential 
velocity does appear to be large.  The system has a proper motion of 
$\mu_\alpha = 4.8 \pm 2.4$ mas~y$^{-1}$ and 
$\mu_\delta = -12.2 \pm 2.3$ mas~y$^{-1}$  
in the Tycho-2 catalogue \citep{hog00}, and 
\citet{rib02} have recently used optical and radio astrometry 
to find an improved estimate of 
$\mu_\alpha = 4.7 \pm 1.1$ mas~y$^{-1}$ and 
$\mu_\delta = -10.6 \pm 1.0$ mas~y$^{-1}$.  
We used the latter measurement together with the systemic 
radial velocity from \citet{mcs01a} (adjusted to the 
velocity offset of the \ion{O}{3} $\lambda 5592$ line that 
presumably forms deep in the photosphere and is less 
affected by expansion in the atmosphere) to find the peculiar 
component of space motion, $v_{\rm pec}$ (following the methods described
in \citet{ber01}), and this quantity is listed in the 
final row of Table~2.   We find that there is satisfying 
agreement between the predicted and observed space velocities,
especially towards the higher mass range.   Note that if
the eccentricity had decreased since the the explosion, 
then the predicted velocities would be lower than the 
observed velocities, which is contrary to the results in 
Table~2.  This strengthens our assumption that little or no
circularization has occurred in LS~5039 since the SN. 

The example of LS~5039 provides a strong confirmation of  
the predictions made about the eccentricity and runaway velocity 
that result from a SN explosion in a binary, and we find 
that both the observed eccentricity and peculiar space velocity 
can be consistently explained by our derived set of pre- and 
post-SN parameters.   On the other hand, the system may be exceptional 
among MXRBs in the huge amount of mass lost in the SN. 
Indeed, LS~5039 appears to be the fastest runaway object 
known among the MXRBs \citep{kap01}.  
The example of LS~5039 hints that there are  
other similar systems with low X-ray luminosity and small 
radial velocity variations, but this class of ``X-ray quiet'' 
SN descendants will be difficult to detect \citep{gar80}. 
  
%%%%%%%%%%%%%%%%%%%%%%%%%%%%%%%%%%%%%%%%%%%%%%%%%%%%%%%%%%%%%%%

\acknowledgments

We thank the KPNO staff, and in particular Diane Harmer
and Daryl Willmarth, for their assistance in making these
observations with the KPNO Coude Feed Telescope.
Institutional support has been provided from the GSU College
of Arts and Sciences and from the Research Program Enhancement
fund of the Board of Regents of the University System of Georgia,
administered through the GSU Office of the Vice President
for Research. 

%%%%%%%%%%%%%%%%%%%%%%%%%%%%%%%%%%%%%%%%%%%%%%%%%%%%%%%%%%%%%%%

% References

%%%%%%%%%%%%%%%%%%%%%%%%%%%%%%%%%%%%%%%%%%%%%%%%%%%%%%%%%%%%%%%

% Figures

\clearpage

% Figure 1
\begin{figure}
%\plotone{f1.eps}
\caption{The average spectrum of LS~5039 after shifting for orbital 
motion and convolution to the resolution of the 1999 spectra.   
The 2000 average spectrum is also plotted with 
the same offsets as the 1998 and 1999 averages
({\it dotted lines}) to emphasize the greater residual 
H$\alpha$ $\lambda 6563$ emission observed in the first two runs.}
\label{fig1}
\end{figure}

% Figure 2
\begin{figure}
%\plotone{f2.eps}
\caption{The predicted X-ray luminosity from wind accretion for 
an assumed O-star mass of $25 M_\odot$ and a range of masses for the 
companion.  The solid line illustrates the result for the average 
mass loss rate calculated for $M_O = 25 M_\odot$ ($\log \dot{M} = -6.3$
for mass loss in $M_\odot$~y$^{-1}$) while the dot-dashed line and 
dashed line show the results for the strongest and weakest mass loss 
rates (from the 1999 and 2000 runs, respectively; $\log \dot{M}$ is 
labelled in each case).  The shaded region 
gives the estimated X-ray luminosity based on a distance of 2.6~kpc 
(the estimated distance for $M_O = 25 M_\odot$) and 
the observed X-ray fluxes reported by \citet{rib99}.
The best match occurs for $M_X=1.9 M_\odot$ in this case.}
\label{fig2}
\end{figure}

% Figure 3
\begin{figure}
%\plotone{f3.eps}
\caption{The mass plane diagram for LS~5039 with the constraints from 
the wind accretion model.  The thick, solid line shows the estimated 
relationship between assumed O-star mass and the mass of the X-ray source. 
The surrounding shaded regions show how the solutions change if we adopt 
the higher mass loss rate from the 1999 observations (lower $M_X$ required 
to obtain the same observed X-ray luminosity) or the lower mass loss rate 
from the 2000 spectra (upper $M_X$ limit).  Lines of constant orbital 
inclination are shown as thin, solid lines.    The dashed line indicates 
the lower limit on $M_X$ established by the lack of X-ray or optical 
eclipses \citep{rib99,cla01}.}
\label{fig3}
\end{figure}

%%%%%%%%%%%%%%%%%%%%%%%%%%%%%%%%%%%%%%%%%%%%%%%%%%%%%%%%%%%%%%%
% Tables

\clearpage

% Table 1 - Sample Stellar Parameter range 

\begin{deluxetable}{lcccc}
\tablewidth{0pc}
\tablecaption{Range in Stellar Parameters\label{tab1}}
\tablehead{
\colhead{Parameter} &
\colhead{$M_O = 10 M_\odot$} &
\colhead{$M_O = 20 M_\odot$} &
\colhead{$M_O = 30 M_\odot$} &
\colhead{$M_O = 40 M_\odot$} }
\startdata
$R_O$ ($R_\odot$)  \dotfill                 & 5.2  & 7.4  & 9.1  & 10.5 \\
$v_\infty$ (km s$^{-1}$) \dotfill           & 2018 & 2400 & 2656 & 2854 \\
$\log \dot{M}$ ($M_\odot$ y$^{-1}$)\dotfill &$-6.7$&$-6.4$&$-6.2$&$-6.1$\\
$M_X$ ($M_\odot$) \dotfill                  & 1.2  & 1.7  & 2.1  & 2.4  \\
$R_O{\rm (Roche)}$ ($R_\odot$) \dotfill     & 8.0  & 10.5 & 12.2 & 13.6 \\
$d$ (kpc) \dotfill                          & 1.7  & 2.4  & 2.9  & 3.3  \\
\enddata
\end{deluxetable}
\clearpage

%%%%%%%%%
% Table 2 - Pre-/Post-SN Parameters

\begin{deluxetable}{lcccc}
\tablewidth{0pc}
\tablecaption{Binary Parameters\label{tab2}}
\tablehead{
\colhead{Parameter} &
\colhead{$M_O = 10 M_\odot$} &
\colhead{$M_O = 20 M_\odot$} &
\colhead{$M_O = 30 M_\odot$} &
\colhead{$M_O = 40 M_\odot$} }
\startdata
$P^{\rm initial}$ (d) \dotfill                       & 1.56   & 1.56   & 1.56   & 1.56  \\
$a^{\rm initial}$ ($R_\odot$) \dotfill               & 14.2   & 17.7   & 20.2   & 22.2  \\
$M_2^{\rm initial}$ ($M_\odot$) \dotfill             &  5.7   & 10.6   & 15.2   & 19.8  \\
$\triangle M_2$ ($M_\odot$) \dotfill                 &  4.6   &  8.9   & 13.2   & 17.4  \\
$R_O$ ($R_\odot$)  \dotfill                          &  5.2   &  7.4   &  9.1   & 10.5  \\
$R_O{\rm (Roche)}^{\rm initial}$ ($R_\odot$)\dotfill &  6.1   &  7.7   &  8.9   &  9.8  \\
$R_2{\rm (Roche)}^{\rm initial}$ ($R_\odot$)\dotfill &  4.7   &  5.8   &  6.5   &  7.1  \\
$v_{\rm sys}$ (predicted) (km s$^{-1}$) \dotfill    &$120\pm10$&$154\pm13$&$178\pm15$&$197\pm17$\\
$v_{\rm pec}$ (observed) (km s$^{-1}$) \dotfill      &$83\pm13$&$118\pm17$&$145\pm21$&$168\pm24$\\
\enddata
\end{deluxetable}
\clearpage

%M_p,post           10           20           30           40
%M_s,post           1.16780      1.68240      2.09040      2.41670
%M_s,pre            5.74660      10.5722      15.2475      19.8075
%P_pre              1.56478
%a_pre              14.2115      17.7291      20.2043      22.1733
%R_p,pre            5.23442      7.40258      9.06627      10.4688
%R_roche,p          6.07719      7.71309      8.86054      9.77545
%R_roche,s          4.71987      5.76717      6.50800      7.09565
%d                  1.66690      2.35730      2.88710      3.33380
%v_run (predicted)  119.687      153.810      177.650      196.666
%v_run (observed)   83.001058    117.99507    145.11764    167.93816
%v_run (errors)     12.765763    17.346962    20.949420    24.017886

%%%%%%%%%%%%%%%%%%%%%%%%%%%%%%%%%%%%%%%%%%%%%%%%%%%%%%%%%%%%%%%
% Figures

\clearpage

\setcounter{figure}{0}

\begin{figure}
\plotone{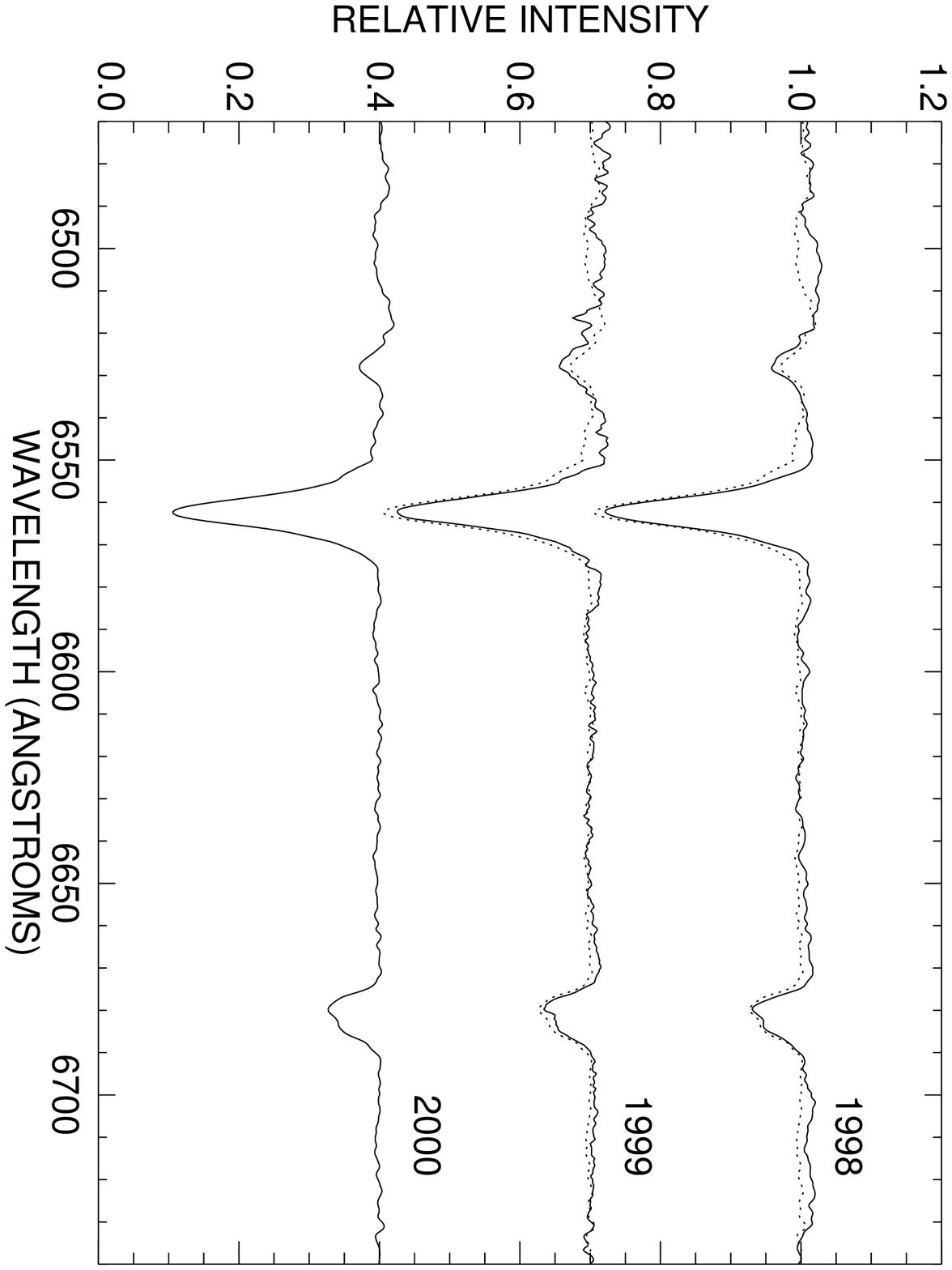}
\caption{}
\end{figure}

\begin{figure}
\plotone{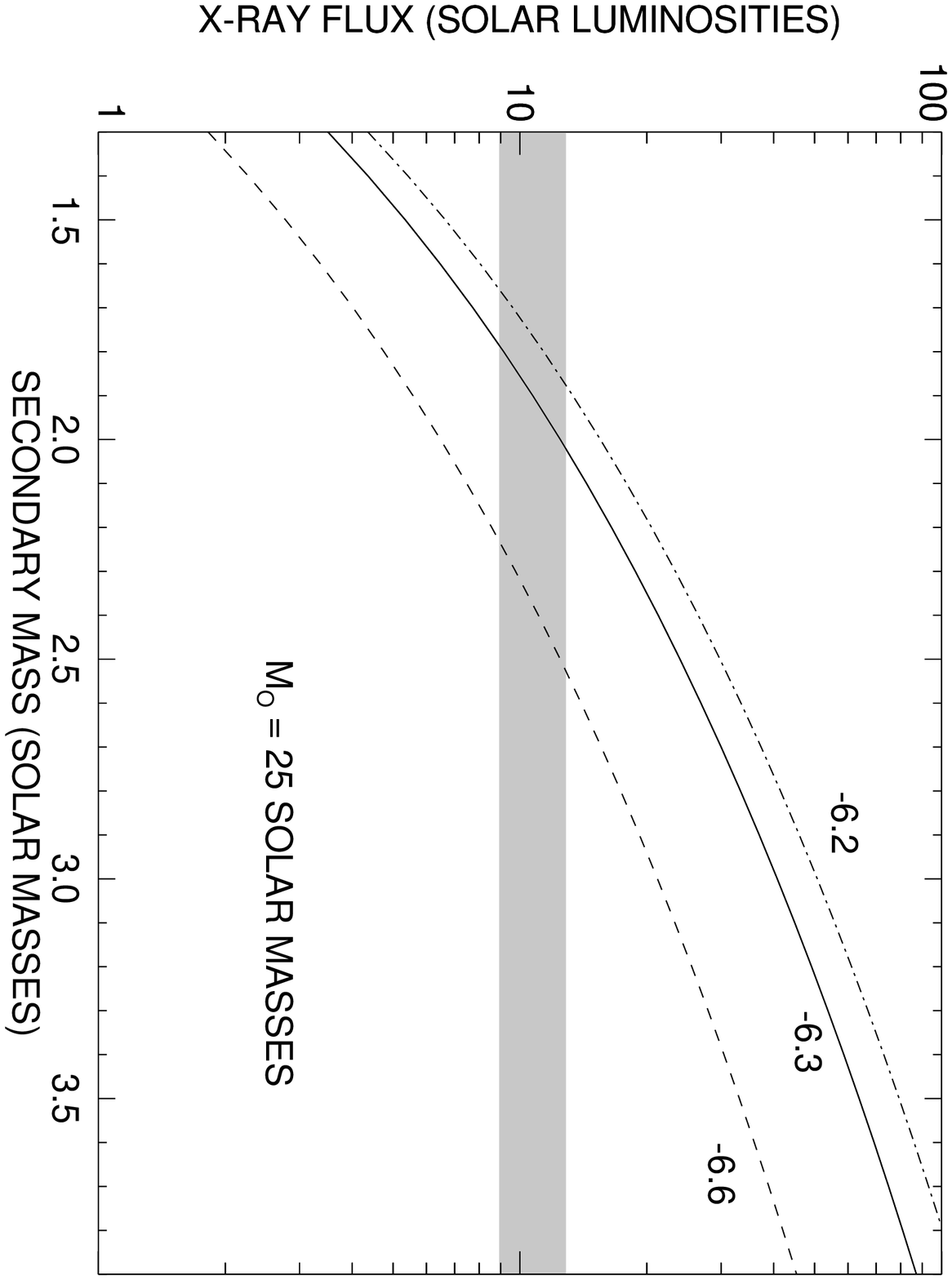}
\caption{}
\end{figure}

\begin{figure}
\plotone{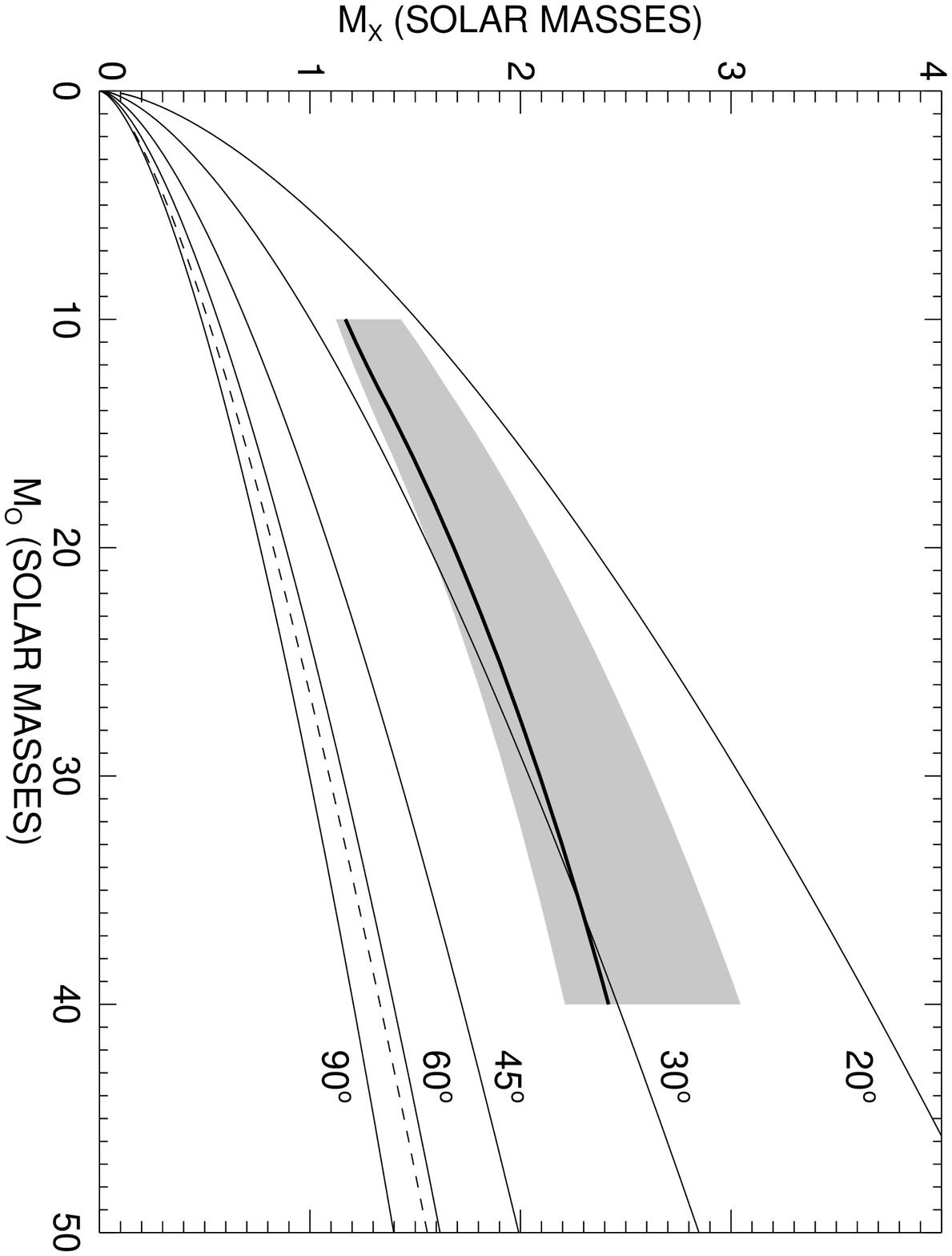}
\caption{}
\end{figure}

%%%%%%%%%%%%%%%%%%%%%%%%%%%%%%%%%%%%%%%%%%%%%%%%%%%%%%%%%%%%%%%

\end{document}